\documentclass[aps,prl,twocolumn,floats,showpacs]{revtex4}
\usepackage{epsfig}
\usepackage{graphicx}
\usepackage{amsmath}

\newcommand{\beq}{\begin{equation}}
\newcommand{\eeq}{\end{equation}}
\newcommand{\bea}{\begin{eqnarray}}
\newcommand{\eea}{\end{eqnarray}}
\newcommand{\eq [1]}{Eq.(\ref{#1})}
\newcommand{\e[1]}{(\ref{#1})}
\newcommand{\eqs [1]} {Eqs.(\ref{#1})}

\begin{document}
\title{Field Theory for the Global
Density of States Distribution Function in Disordered Conductors}
\author{
%\footnote{Present address: }
V.I.Yudson}

\affiliation{Institute for Spectroscopy, Russian Academy of Sciences, Troitsk, Moscow
region, 142190 Russia}

%\date{October 18, 2004}
%\today}

\begin{abstract}
A field-theoretical representation is suggested for the electron global density of
states distribution function $\mathcal{P}(\nu)$ in extended disordered conductors.
This opens a way to study the complete statistics of fluctuations. The approach is
based on a functional integration over bi-local functions
$\Psi_{\mathbf{r}_1,\mathbf{r}_2}$ instead of the integration over local functions in
the usual functional representation for moments of physical quantities. The formalism
allows one to perform the disorder averaging and to derive an analog of the usual
nonlinear $\sigma$-model - a ``slow" functional of a supermatrix field
$Q_{\mathbf{r}_1,\mathbf{r}_2}(\mathbf{r}) \sim \Psi_{\mathbf{r},\mathbf{r}_1} \circ
\bar{\Psi}_{\mathbf{r}_2,\mathbf{r}}$. As an application of the formalism, the
long-tail asymptotics of $\mathcal{P}(\nu)$ is derived.
\end{abstract}

\maketitle

A powerful theoretical tool for studying localization phenomena in disordered media
is based on the functional integral representation of the Green's functions. This
representation is the starting point for the derivation of an effective slow action -
the nonlinear $\sigma$-model \cite{sigma,SUSY}. The renormalization group (RG)
analysis of this model describes the scaling of the \emph{averaged} conductance $<g>$
with the system size $L$ in accordance with the one-parameter scaling hypothesis
\cite{Abrahams79}. Since the discovery of mesoscopic fluctuations (see review
\cite{Meso}) the importance of studying distribution functions has been realized. For
systems of dimension $d \leq 1$, distribution functions of some physical parameters
can be found with the use of a ``zero-dimensional" supersymmetric $\sigma$-model or
transfer-matrix formalism (see references in the books \cite{SUSY,Meso}). For
extended systems ($d > 1$), distribution functions of some \emph{\textit{local}}
quantities - like eigenfunction amplitudes or the local density of states (DOS) -
have been obtained \cite{Local} by means of the saddle-point approach \cite{MK} to
the supersymmetric $\sigma$-model. However, no suitable field-theoretical
representation for the \emph{global} DOS and conductance distribution functions,
$\mathcal{P}(\nu)$ and $\mathcal{P}(g)$, in systems of dimension $2 \leq d \leq 4$
has been known up to now; the information about these functions has been extracted
from calculation of all the moments $<\nu^n>$ and $<g^n>$ \cite{AKL}. A field theory
for distribution functions is certainly in demand.

In the present paper we develop such a theory for $\mathcal{P}(\nu)$ and derive a
slow effective functional \eqs[F0]-\e[Fsnu] that determines the characteristic
function $\mathcal{P}_{\nu}(s)$ :
\begin{eqnarray}\label{Ps}
\mathcal{P}_{\nu}(s) \equiv <\exp{[- s \nu/\bar{\nu}]}> = \int
\mathcal{D}[Q]\,\exp{[F_0 + F_s]}\, ,
\end{eqnarray}
where $<\ldots>$ denotes the averaging over the disorder potential $U$; $\bar{\nu} =
<\nu>$, and the global DOS, $\nu$, is determined for a particular realization of the
disorder by
\begin{eqnarray}\label{nu}
\nu(E) = i/(2\pi L^d)\mathrm{Tr}\{G^R(E) - G^A(E)\} \, .
\end{eqnarray}
Here $G^{R(A)}(E) = [E - H \pm i\delta]^{-1}$ are exact retarded and advanced
electron Green's functions for the Hamiltonian $H = H_0 + U$; the trace $\mathrm{Tr}$
is taken over space arguments.

For the unitary ensemble (spinless electrons in the presence of a time-inverse
symmetry breaking magnetic field) the ``free" action $F_0$ is given by
\begin{eqnarray}\label{F0}
F_0 = \frac{\pi \bar{\nu}}{4}\int{d\mathbf{r}\,\mathrm{Str}\{D(\nabla
Q(\mathbf{r}))^2 - 2\delta\Lambda_z Q(\mathbf{r})\}} \, ;
\end{eqnarray}
$D = v_Fl/d$ is the diffusion coefficient; $l>>p^{-1}_F$ is the mean free path, $p_F$
and $v_F$ are the Fermi momentum and velocity; $\delta$ is a dephasing rate. The
``source" action $F_s$ is
\begin{eqnarray}\label{Fsnu}
F_s = - \tilde{s}\int{d\mathbf{r}\,\mathrm{Str}\{O^{(\nu)}(\mathbf{r})\Lambda_z k
Q(\mathbf{r})\}} \, ,
\end{eqnarray}
where $\tilde{s} = s\tau\Delta/(2\pi)$, $\tau = l/v_F$, and $\Delta =
1/(\bar{\nu}L^d)$.

In contrast to the usual $\sigma$-models, supermatrices $Q$ in \eqs [Ps],\e[F0], and
\e[Fsnu] depend on \emph{three} spacial arguments: $Q(\mathbf{r}) = \{Q_{\mathbf{r}_1
\mathbf{r}_2}(\mathbf{r})\}$; the dependence on $\mathbf{r}$ is smooth. The symbol
``$\mathrm{Str}$" denotes a combination of the usual supertrace operation
$\mathrm{str}\{M\} = \mathrm{tr}\{M^{bb}\} - \mathrm{tr}\{M^{ff}\}$ ($b$ and $f $
refer to boson and fermion components) and the summation over the (lower) spacial
arguments: $\mathrm{Str}\{M\} \equiv \sum_{\mathbf{r}_1}\mathrm{str}\{M_{\mathbf{r}_1
\mathbf{r}_1}\}$. The summation runs over sites of an auxiliary lattice of spacing
$a$; in the continuum limit ($a \rightarrow 0$) the summation is replaced for
$\int{d\mathbf{r}_1/a^d}$.

In \eqs [F0]-\e[Fsnu], $Q$ is a $4 \times 4$ matrix in the kinetic R-A (retarded -
advanced) space and in the b-f superspace; $\Lambda_i$ ($i = x,y,z)$ are the Pauli
matrices acting in the R-A space; $k = \mathrm{diag}(1,-1)$ is a supermatrix in the
b-f superspace. The supermatrix $Q$ obeys the constraint
\begin{eqnarray}\label{Q2}
Q^2(\mathbf{r}) = I \, ,
\end{eqnarray}
where $I$ is the unity in both the intrinsic matrix space and the coordinate space.
The operator $O^{(\nu)}(\mathbf{r})$ is defined as
\begin{eqnarray}\label{Onu}
O^{(\nu)}_{\mathbf{r}_1\mathbf{r}_2}(\mathbf{r}) = C_{\mathbf{r}_2 \mathbf{r}_1}
\delta[\mathbf{r} - ({\mathbf{r}_1 + \mathbf{r}_2})/2] \, ;
\end{eqnarray}
$C = [1 + 4\tau^2\zeta^2]^{-1}$, where $\zeta$ is the operator of kinetic energy
counted from the Fermi energy $E_F$.

The slow functionals \eqs[F0] and \e[Fsnu] allow one to study the complete statistics
of fluctuations $\nu$ in disordered conductors. Earlier, the way of studying
$\mathcal{P}(g)$ and $\mathcal{P}(\nu)$ was to calculate all the moments $<g^n>$ and
$<\nu^n>$. This formidable problem has been attacked in \cite{AKL} with the use of an
\emph{extended} $\sigma$-model. The matrix variable $Q^{i,j}(r)$ of this model has
additional indices $(i,j = 1, \ldots, n)$ to account for coupling (at the averaging
over the disorder) of primary fields $\bar{\Psi}^i(\mathbf{r})$ and
$\Psi^i(\mathbf{r})$ used in the standard functional integral representation of each
term ($i = 1, \ldots, n)$ of a product $g^n$ or $\nu^n$. The action of the
\emph{extended} $\sigma$-model includes all the terms of the expansion in powers of
$l\nabla Q$ and $\omega\tau Q$, while the usual $\sigma$-model corresponds to the
``hydrodynamical" approximation, i.e. to the lowest non-zero terms $\sim (\nabla
Q)^2$ and $\sim \omega Q$; $\omega$ is the frequency of an external field. The formal
reason of importance of short distances has been revealed \cite{KLY} by the RG
analysis of the extended $\sigma$-model: vertices with $2n$ gradients have a positive
anomalous dimension $\sim (n^2 - n)$. This anomaly leads to an ``explosion" of high
cumulant moments, so that the distribution functions $\mathcal{P}(g)$ and
$\mathcal{P}(\nu)$ possess logarithmically normal (LN) tails \cite{AKL}:
\begin{eqnarray}\label{LN}
\mathcal{P}(\nu) \propto \exp{\left[-(1/4u)\ln^2{\left[\delta \nu /
\bar{\nu}\tau\Delta\right]}\right]} \,\, ,
\end{eqnarray}
where $\delta \nu = \nu - \bar{\nu}$ and $u = \ln{(\sigma_0/\sigma)}$ ($\sigma_0$ is
a bare value of the conductivity $\sigma$ for the \emph{orthogonal} ensemble). The
same mechanism leads also to the long-time asymptotics of the averaged conductance
$<g(t)>$ \cite{AKL88}; it has been interpreted in terms of the LN distribution of the
current relaxation times $t_{\phi}$:
\begin{eqnarray}\label{LNt}
\mathcal{P}(t_{\phi}) \propto \exp{\left[-(1/4u)\ln^2{(t_{\phi}/{\tau})}\right]}\,\,.
\end{eqnarray}
Assuming that the asymptotics \eq [LN] is caused by a \textit{single}
``quasi-localized" state of a small energy width $\sim 1/t_{\phi}$ with the
corresponding contribution $\delta\nu \sim t_{\phi}$ to the global DOS, one may
obtain \eq [LN] directly from \eq [LNt] \cite{K92,Mirlin97}. On the other hand, no
similar reasoning connecting \eq [LNt] with the LN asymptotics of the conductance
distribution function has been proposed yet.

The distributions \eq [LN] and \eq [LNt] have been obtained \cite{AKL,AKL88} with the
use of the sophisticated RG analysis \cite{KLY} restricted formally to $d=2+\epsilon$
case. Later on, an alternative approach to study the asymptotics of $<g(t)>$ has been
suggested in \cite{MK}. Based on the saddle-point analysis of the usual
(\emph{non-extended}) supersymmetric $\sigma$-model for the \emph{averaged}
conductance, this approach reproduces \eq [LNt] at d=2 and gives results for other
dimensions \cite{MK,Mirlin95}. The problem of short distances in this approach was
treated by introducing a cutoff $r_* \sim l$ \cite{MKbal}.

It is tempting to confirm in a similar way the RG result \eq [LN] for the global DOS
(as well as for the conductance) distribution function. However, this task has
remained unfeasible because of the absence of a working field-theoretical
representation for $\mathcal{P}(\nu)$ (and $\mathcal{P}(g)$). Below we shall achieve
this goal with the use of the developed theory, \eqs [F0] and \e[Fsnu].

For completeness, we briefly comment on a seemingly simple but hardly useful
representation for $\mathcal{P}(\nu)$. It is based on the integration in \eq [Ps]
over supermatrices $Q^{i,j}(\mathbf{r})$ which have only one spacial argument but an
infinite number of colors: $1 \leq i,j \leq \mathcal{N}$; $\mathcal{N} \rightarrow
\infty$. The free action $F_0$ is still given by \eq [F0] (with $\mathrm{Str}
\rightarrow \mathrm{str}$), and $F_s = -(s/4\mathcal{N})\int \mathrm{str}\{\Lambda_z
k Q(\mathbf{r})\}d\mathbf{r}/L^d$. However, the advantage of the supersymmetry - a
finite rank of the Q-matrix - is lost and it is not clear how to treat
non-perturbative regimes; e.g. there is no apparent way to find solutions to the
saddle-point equation $\mathbf{\nabla}(Q\nabla Q) =
[s\Delta/(2\pi\mathcal{N}D)][\Lambda_z k , Q(\mathbf{r})]$ and the corresponding
action, which would behave properly in the limit $\mathcal{N} \rightarrow \infty$.

The formalism leading to the effective field theory of interest, \eqs [F0] and
\e[Fsnu], is based on the following representation of $\mathcal{P}_{\nu}(s)$ in the
form of a functional integral over \emph{bi-local} primary superfields
$\bar{\Psi}_{\mathbf{r}_1\mathbf{r}_2}$ and $\Psi_{\mathbf{r}_1\mathbf{r}_2}$:
\begin{eqnarray}\label{Pnu}
\mathcal{P}_{\nu}(s) = \int
\mathcal{D}[\bar{\Psi},\Psi]<e^{i[\bar{\Psi}\mathcal{M}\Psi -
\tilde{s}(\bar{\mathcal{B}}\Psi + \bar{\Psi} \mathcal{B})]}> .
\end{eqnarray}
Here the superfield $\Psi_{\mathbf{r}_1\mathbf{r}_2}$ is defined as a column of
complex ($S^{R(A)}$) and Grassmann ($\xi^{R(A)}$) fields $\Psi = (S^R, S^A; \xi^{R},
\xi^{A})^t$ and $\bar{\Psi} = \Psi^+ \hat{L}$, where the matrix $\hat{L} =
\mathrm{diag}(L_b, L_f) = \mathrm{diag}(\Lambda_z, I)$; the transposition ``$t$" does
not touch space variables, while the Hermitian conjugation is defined with their
interchange. An explicit form of the scalar product is $\bar{\Psi}\mathcal{M}\Psi =
\bar{\Psi}_{\mathbf{r}_2\mathbf{r}_1} \mathcal{M}_{\mathbf{r}_1\mathbf{r}_2 ;
\mathbf{r}_3\mathbf{r}_4} \Psi_{\mathbf{r}_4\mathbf{r}_3}$ (with the summation over
repeated indices), where
\begin{eqnarray}\label{M}
\mathcal{M}_{\mathbf{r}_1\mathbf{r}_2 ; \mathbf{r}_3\mathbf{r}_4} =
\delta_{\mathbf{r}_2\mathbf{r}_3}\tau[E - H +
i\delta\Lambda_z/2]_{\mathbf{r}_1\mathbf{r}_4} \, .
\end{eqnarray}
In \eq [Pnu], $\mathbf{\mathcal{B}}^{\alpha}_{\mathbf{r}_1\mathbf{r}_2} =
B\delta_{\mathbf{r}_1\mathbf{r}_2}\delta_{\alpha b}$ (and similarly for
$\bar{\mathcal{B}}$) has only the boson component; the constant vectors $B$ and
$\bar{B}$ in the R-A space are specified by their direct product $B \circ \bar{B}
\equiv \Omega = \Lambda_z + i\Lambda_y$. The choice of the matrix $\Omega$ is
dictated by the requirements of the factorizability of $\Omega$ and convergence of
integrals in the slow functional (details will be presented in an extended
publication). We choose $\bar{B} = \mathrm{diag}(1,1)$ and $B = \mathrm{diag}(1,
-1)^t$. Integrating over $\bar{\Psi}$,$\Psi$ in \eq [Pnu], we find:
\begin{eqnarray}\label{TLnu}
\mathcal{P}_{\nu}(s)=<\exp[-\mathrm{Str}\mathrm{ln}\{\mathcal{M}\}-
i\tilde{s}\mathrm{Str}\{\mathcal{M}^{-1}\Omega\Pi_b\Pi\}]>  .
\end{eqnarray}
Here $\Pi_b = (1 + k)/2$ is the projector on the boson sector of the b-f superspace;
$\Pi_{\mathbf{r}_1\mathbf{r}_2 ; \mathbf{r}_3\mathbf{r}_4} =
\delta_{\mathbf{r}_1\mathbf{r}_2}\delta_{\mathbf{r}_3\mathbf{r}_4}$. The first term
in the exponent vanishes due to the supersymmetry (matrix $\mathcal{M}$ is diagonal
in the b-f superspace) while the second one equals $-s\nu/\bar{\nu}$, thus providing
the desired equality $\mathcal{P}_{\nu}(s) = <\exp[- s\nu/\bar{\nu}]>$.

The primary representation \eq [Pnu] allows one to average the term
$\exp{[-i\sum_{\mathbf{r}, \mathbf{r}'}\bar{\Psi}_{\mathbf{r}',
\mathbf{r}}U_{\mathbf{r}}\Psi_{\mathbf{r}, \mathbf{r}'}]}$ over the Gaussian disorder
potential $U$ and to use the Hubbard-Stratonovich (HS) decoupling of the term
$(\bar{\Psi}_{\mathbf{r}_1\mathbf{r}}
\Psi_{\mathbf{r}\mathbf{r}_1})(\bar{\Psi}_{\mathbf{r}_2\mathbf{r}}
\Psi_{\mathbf{r}\mathbf{r}_2})$ with an auxiliary matrix field
$Q_{\mathbf{r}_1\mathbf{r}_2}(\mathbf{r})$ of the usual symmetry $Q^\dag =
\hat{L}Q\hat{L}$ (see \cite{SUSY}). After this transformation, the bilinear over
$\bar{\Psi}$ and $\Psi$ part of the action \eq [Pnu] will be modified in the
following way: $\mathcal{M} \rightarrow \mathcal{M}(\mathcal{Q}) = \mathcal{M}_0 +
i\mathcal{Q}/2$, where $\mathcal{M}_0$ is given by \eq [M] with $H=H_0$;
$\mathcal{Q}_{\mathbf{r}_1\mathbf{r}_2 ; \mathbf{r}_3\mathbf{r}_4} \equiv
\delta_{\mathbf{r}_1\mathbf{r}_4}Q_{\mathbf{r}_2\mathbf{r}_3}(\mathbf{r}_1)$.
Integration over $\bar{\Psi}$, $\Psi$ leads to \eq [TLnu] where $\mathcal{M}$ is
replaced by $\mathcal{M}(\mathcal{Q})$ and the averaging goes over the Gaussian
action for the HS field $Q$. A (primary) saddle-point approach to the modified \eq
[TLnu] reveals a manifold \eq [Q2] where the action has a degenerate extremum
perturbed by the $\delta\Lambda_z/2$ term and by the ``source" term ($\propto
\tilde{s}$). Restricting to only the "transverse" variations of $Q$ (obeying \eq
[Q2]) and keeping only the lowest powers of $\nabla Q$ (this hydrodynamical
approximation corresponds to a non-extended $\sigma$-model), we obtain \eq [Ps] for
$\mathcal{P}_{\nu}(s)$ with the ``free" action $F_0$ \eq [F0] and the "source" action
$F_s$ given by
\begin{eqnarray}\label{Fsnu1}
F_s = - \tilde{s}\int{d\mathbf{r}\,\mathrm{Str}\{O^{(\nu)}(\mathbf{r}) (\Lambda_z +
i\Lambda_y)Q(\mathbf{r})(1 + k)\}} \, .
\end{eqnarray}
With the use of the parametrization \cite{SUSY}: $Q = \Lambda_z(I + iP)(I - iP)^{-1}$
(with $\Lambda_zP = - P\Lambda_z$), we find that the ``$\Lambda_z$-part" of the
action \eq [Fsnu1] is even in $P$, while the ``$\Lambda_y$-part" is odd. The diagrams
generated by the $\Lambda_y$-part of \eq [Fsnu1] are similar to those which would be
caused by higher terms of the gradient expansion of
$\mathrm{Str}\,\mathrm{ln}\{\ldots\}$, if we allowed for effects of the extended
$\sigma$-model. In the considered hydrodynamic approximation all such terms should be
ignored. Also, having in mind the assumed smallness of the source term, we shall keep
in \eq [Fsnu1] only the part ($\propto k$) which breaks down the symmetry between the
boson and fermion components of the $Q$-action (if not this part, the integral \eq
[Ps] would be just unity for any $s$). Omitting in \eq [Fsnu1] the unity term of the
projector $(1+k)$ and the ``$\Lambda_y$-part", we arrive at the announced result, \eq
[Fsnu].

To check the agreement with the usual diagrammatic technique, we treat $F_s$ as a
perturbation and calculate $\mathrm{ln}[\mathcal{P}_{\nu}(s)]$ as a series in powers
of $s$. This calculation can be performed with the use of the mentioned
parametrization of $Q$ in terms of the non-constrained field $P$. The leading
contribution to the term of a given order in $s$ is made by diagrams with a minimal
number of loops formed by lines associated with propagators (``diffusons") $<PP>$ of
the field $P$. For instance, for the the first and second order terms of
$\mathrm{ln}[\mathcal{P}_{\nu}(s)]$ we find $-s$ and $s^2\Delta^2/(2\pi)^2
\sum_{\mathbf{q} \neq 0}[D\mathbf{q}^2]^{-2}$, respectively. From here we obtain the
average $<\nu/\bar{\nu}> = 1$ and the variance $<(\delta\nu/\bar{\nu})^2> =
\Delta^2/(2\pi^2) \sum_{\mathbf{q} \neq 0}[D\mathbf{q}^2]^{-2}$ coinciding with the
results of the diagrammatic technique \cite{AS86}. The latter expression for the
variance $<(\delta\nu/\bar{\nu})^2>$ is determined by one-loop diagrams. Higher
cumulant moments of $\delta\nu$ are determined by diagrams with higher number of
loops \cite{AKL}. Using \eqs [F0] and \e[Fsnu], we have checked a non-trivial
``accidental" cancellation of two-loop contributions to the third-order cumulant
$<(\delta\nu/\bar{\nu})^3>_c$, so the latter is determined by three-loop diagrams, in
accordance with \cite{AKL}.

Most interesting applications of the developed formalism, \eqs [F0] and \e[Fsnu], are
expected in the \emph{non-perturbative} regime. In the present paper we implement the
idea of the approach \cite{MK} and consider non-trivial ``secondary" saddle-point
configurations of the action $F = F_0 + F_s$. Looking for an extremum of $F$ with
respect to variations of the field $Q$ (obeying \eq [Q2]): $Q \rightarrow Q + \delta
Q$, $\delta Q = [Q, \, \epsilon]$, we obtain the following ``secondary" saddle-point
equation:
\begin{eqnarray}\label{SP1}
\mathbf{\nabla}[Q(\mathbf{r})\mathbf{\nabla}Q(\mathbf{r})] =
\frac{\tilde{s}}{\pi\bar{\nu}D}[\Lambda_z k O^{(\nu)}(\mathbf{r}), Q(\mathbf{r})] \,
.
\end{eqnarray}
Below we restrict the analysis to a particular sub-manifold $\mathcal{E}_0$ of the
saddle-point manifold $\mathcal{E}$: matrices $Q_{\mathbf{r}_1
\mathbf{r}_2}(\mathbf{r}) \in \mathcal{E}_0$ are characterized by a smooth dependence
on the ``center-of-mass" coordinate $(\mathbf{r}_1 + \mathbf{r}_2)/2$ (i.e.,
$|Ql\nabla'Q| \ll 1$) and by a small difference $|\mathbf{r}_1 - \mathbf{r}_2| \sim
l$.

It is convenient to re-write \eqs [F0] and \e[Fsnu] in the Wigner representation over
the spacial indices: $A_{\mathbf{r}_1\mathbf{r}_2} =
(a^d/L^{d})\sum_{\mathbf{p}}A(\mathbf{r}'; \mathbf{p})\exp[i\mathbf{p}(\mathbf{r}_1 -
\mathbf{r}_2)]$; $\mathbf{r}' = (\mathbf{r}_1 + \mathbf{r}_2)/2$. Omitting for
brevity the $\delta$-term in \eq [F0], we have:
\begin{eqnarray}\label{F0W}
F_0 = \frac{\pi\bar{\nu}}{4} \int
\frac{d\mathbf{r}d\mathbf{r}'}{L^d}\sum_{\mathbf{p}}\,
\mathrm{str}\{D[\nabla_{\mathbf{r}} Q(\mathbf{r}; \mathbf{r}'; \mathbf{p})]^2\} \,
,\\
F_s = - \tilde{s}\int{ \frac{d\mathbf{r}}{L^d}\sum_{\mathbf{p}} C_p \,
\mathrm{str}\{\Lambda_z k Q(\mathbf{r}; \mathbf{r}; \mathbf{p})\}} \, , \label{FsnuW}
\end{eqnarray}
where $C_p = [1 + 4\tau^2\zeta^2_p]^{-1}$; $\zeta_p = v_F(p - p_F)$. In the
considered hydrodynamical approximation (non-extended $\sigma$-model) we may re-write
the constraint \eq [Q2] in the form:
\begin{eqnarray}\label{Q2W}
Q^2(\mathbf{r}; \mathbf{r}'; \mathbf{p}) = I
\end{eqnarray}
for any value of arguments. The transformation from \eqs [F0] and \e[Fsnu] to \eqs
[F0W] and \e[FsnuW] is exact, while the form \eq [Q2W] of the original constraint \eq
[Q2] is appropriate only for the sub-manifold $\mathcal{E}_0$. On this sub-manifold,
the saddle-point equation \e[SP1] looks like
\begin{eqnarray}\label{SP2}
&&\mathbf{\nabla}_{\mathbf{r}}[Q(\mathbf{r},\mathbf{r}';
\mathbf{p})\mathbf{\nabla}_{\mathbf{r}}Q(\mathbf{r},\mathbf{r}'; \mathbf{p})] =
\nonumber \\
&& \tilde{s}C_p/(\pi\bar{\nu}D) [\Lambda_z k, Q(\mathbf{r}',\mathbf{r}'; \mathbf{p})]
\delta(\mathbf{r} - \mathbf{r}') \, .
\end{eqnarray}
Similar to \cite{MK}, we take into account only the non-compact boson degree of
freedom (the angle $\theta$; $0 < \theta < \infty$) and parametrize the boson sector
of $Q$ in the form: $Q^{RR} = -Q^{AA} = \cosh{\theta}$, $Q^{AR} = - Q^{RA} =
\sinh{\theta}$. In this approximation, the distribution function $\mathcal{P}(\nu)$,
determined from \eq [Ps] by the inverse transformation, reads:
\begin{eqnarray}\label{P(nu)}
\mathcal{P}(\nu) = \int d s/(2\pi i)d\theta \exp(s\delta\nu/\bar{\nu} -
F_{b}[\theta]) \, ,
\end{eqnarray}
where the $s$-integration goes along a contour parallel to the imaginary axis and the
boson action $F_{b}$ is
\begin{eqnarray}\label{Fbos}
&&F_{b}[\theta] = \frac{\pi\bar{\nu}D}{2}\sum_{\mathbf{p}}\int
\frac{d\mathbf{r}d\mathbf{r}'}{L^d}[\mathbf{\nabla}_{\mathbf{r}}\theta(\mathbf{r};
\mathbf{r}';
\mathbf{p})]^2 \nonumber \\
&+& 2\tilde{s}\sum_{\mathbf{p}}C_{p}\int \frac{d\mathbf{r}}{L^d}\,
[\cosh{\theta(\mathbf{r}; \mathbf{r}; \mathbf{p})} - 1] \, .
\end{eqnarray}
Integration over $s$ in \eq [P(nu)] leads to the constraint:
\begin{eqnarray}\label{Constr}
\frac{\delta\nu}{\bar{\nu}} = \frac{\tau\Delta}{\pi}\sum_{\mathbf{p}}C_{p}\int
 \frac{d\mathbf{r}}{L^d}\, [\cosh{\theta(\mathbf{r}; \mathbf{r};
\mathbf{p})} - 1] \, .
\end{eqnarray}
The saddle point equation for the action $F_{b}$ \eq [Fbos] is
\begin{eqnarray}\label{SPnu}
\nabla^2_{(\mathbf{r})}\theta(\mathbf{r}; \mathbf{r}'; \mathbf{p}) -
\kappa_p\sinh\theta(\mathbf{r}'; \mathbf{r}'; \mathbf{p}) \delta(\mathbf{r}-
\mathbf{r}') = 0 \, ,
\end{eqnarray}
where $\kappa_p = 2\tilde{s}C_p/(\pi \bar{\nu} D)$. This equation differs
substantially from that derived in \cite{MK} for the usual $\sigma$-model. \eq [SPnu]
is easy to solve. We find:
\begin{eqnarray}\label{theta}
\theta(\mathbf{r}; \mathbf{r}'; \mathbf{p}) = - \kappa_p \mathcal{D}(\mathbf{r};
\mathbf{r}')\sinh\theta(\mathbf{r}'; \mathbf{r}'; \mathbf{p}) \, ,
\end{eqnarray}
where $-\mathcal{D}(\mathbf{r}; \mathbf{r}')$ is the Green's function of the Laplace
operator with the boundary conditions: $\mathcal{D} = 0$ on contacts and
$\mathbf{\nabla}_\mathbf{n}\mathcal{D} = 0$ elsewhere. The value of
$\theta(\mathbf{r}; \mathbf{r}; \mathbf{p})$ is determined by the self-consistency
condition for \eq [theta] at $\mathbf{r} = \mathbf{r}'$ \cite{foot1}. One of
solutions to this equation is trivial: $\theta(\mathbf{r}; \mathbf{r}; \mathbf{p})=
0$.  Assuming $\kappa_p \ll 1$, we find with the logarithmic accuracy a non-trivial
solution: $\theta \mathrm{sign}[\mathrm{Re}(\theta)] \approx
\ln[-1/(\kappa_p\mathcal{D}(\mathbf{r}; \mathbf{r}))]$. To minimize the saddle-point
action, we take into account the non-trivial solution for only one of possible
$\mathbf{p}$-channels (and put $|\mathbf{p}| = p_F$) - this is similar to the
restriction to only one ``pre-localized state" in \cite{MK}. Finding the quantity
$\tilde{s}$ from \eq [Constr], we obtain $\theta(\mathbf{r}; \mathbf{r};
|\mathbf{p}|=p_F) \simeq \ln[\delta\nu/(\bar{\nu}\tau\Delta)]$. Using \eqs [theta]
and \e[Fbos], we get finally for $\mathcal{P}(\nu)$, \eq [P(nu)]:
\begin{eqnarray}\label{LNnu}
\mathcal{P}(\nu) \propto \exp\left(-[\pi\bar{\nu}D/(2\tilde{\mathcal{D}})]
\ln^2\left[\delta\nu/(\bar{\nu}\tau\Delta)\right]\right) \, ,
\end{eqnarray}
where $1/\tilde{\mathcal{D}} = L^{-d} \int \, d\mathbf{r}/\mathcal{D}(\mathbf{r};
\mathbf{r})$. The ``anomalous'' saddle-point contribution \eq [LNnu] to
$\mathcal{P}(\nu)$ has been derived under the condition $\delta\nu/\bar{\nu} \gg
\tau\Delta$ (large $\theta(\mathbf{r}; \mathbf{r})$). This contribution dominates
over the ordinary Gaussian background $\mathcal{P}(\nu) \propto
\exp\left[-(\delta\nu)^2/(2<(\delta\nu)^2>)\right]$ at $\delta\nu/\bar{\nu} \gg
\bar{g}^{-1/2}(l/L)^{d/2 - 1}$ (we define $\bar{g}$ as $\bar{g} = 2\pi E_c/\Delta$).
For $d=2$, $\tilde{\mathcal{D}} \simeq (2\pi)^{-1}\ln(L/l)$ and we have for \eq
[LNnu]:
\begin{eqnarray}\label{LNnu2}
\mathcal{P}(\nu) \propto \exp\left(-\frac{\pi\bar{g}}{2\ln(L/l)}
\ln^2\left[\frac{\delta\nu}{\bar{\nu}\tau\Delta}\right]\right) \, .
\end{eqnarray}
\eq [LNnu2] coincides with \eq [LN] (up to the factor 2 resulting from the difference
between the orthogonal and unitary ensembles) and with the result of the physical
reasoning \cite{K92,Mirlin97} based on \eq [LNt], but it has weaker restrictions on
its validity range. However, at d=3 there is no agreement between \eq [LNnu] and the
result \cite{Mirlin97}, based on a version of \eq [LNt] for d=3 space \cite{MK}.
Also, there is some discrepancy (although not too dramatic) in the
quasi-one-dimensional case: in \cite{Mirlin97}, the logarithm's argument contains an
extra factor $\bar{g}\tau\Delta \sim E_c\tau$. A reason of the discrepancies at $d
\neq 2$ is not clear yet. It may be a sign of importance of fluctuations and
interaction of boson and fermion modes, ignored in the simplified saddle-point
analysis of \cite{MK} and in the present paper (fluctuations can also restrict the
possibility of dealing with only $\mathcal{E}_0$ sub-manifold of the saddle-point
manifold $\mathcal{E}$). Note that the RG results, \eqs [LN] and \e[LNt], refer
formally to only d=2.

To conclude, an exact functional integral representation, \eq [Pnu], is suggested for
the \emph{global} electron DOS distribution function $\mathcal{P}(\nu)$. This
representation allows one to perform the disorder averaging and to derive a ``slow"
functional, \eqs [F0] and \e[Fsnu]. The formalism reproduces the diagrammatic results
for moments of DOS and provides also a possibility to study non-perturbative regimes.
In particular, the logarithmically normal asymptotics of $\mathcal{P}(\nu)$ has been
derived. With some complication of the functional integrals, the approach can be
implemented also for the conductance distribution function $\mathcal{P}(g)$
\cite{Ytbp}.
\begin{acknowledgments}
I am grateful to B.L.Altshuler, I.V.Lerner, A.D.Mirlin, I.E.Smolyarenko, and
especially V.E.Kravtsov for many helpful discussions. The research was supported by
RFFI (grant 03-02-17285) and ``Nanostructures" program of Russian Academy of
Sciences. I acknowledge also the hospitality of the Abdus Salam International Center
for Theoretical Physics where a part of the work was done.
\end{acknowledgments}


\begin{thebibliography}{99}
\bibitem{foot1} Like in Refs. \cite{MK,Mirlin95}, to regularize the continual
theory and to provide the assumed smooth spacial dependence of $\theta(\mathbf{r};
\mathbf{r}'; \mathbf{p})$, we require $|\mathbf{r} - \mathbf{r}'| \geq r_* \gtrsim
l$, $|l\nabla\theta|_ {|\mathbf{r} - \mathbf{r}'| = r_*} \sim 1$, and define
$\mathcal{D}(\mathbf{r},\mathbf{r}) \equiv \mathcal{D}(\mathbf{r},\mathbf{r}')|_
{|\mathbf{r} - \mathbf{r}'| = r_*}$.
\end{thebibliography}
\end{document}